\documentclass[11pt]{article}
\usepackage{amssymb}
\usepackage{amsmath}
\usepackage{amscd}
\usepackage{latexsym}
\usepackage{bbm}

\def\a{\alpha}
\def\b{\beta}
\def\ga{\gamma}
\def\la{\lambda}
\def\ga{\gamma}
\def\de{\delta}
\def\eps{\epsilon}
\def\ve{\varepsilon}
\def\vk{\varkappa}
\def\Si{\Sigma}
\def\p{\phi}
\def\vp{\varphi}
\def\th{\theta}
\def\O{\Omega}

\newcommand{\R}{\mathbb R}

\newcommand{\Gcal}{{\cal G}}
\newcommand{\Acal}{{\cal A}}

\newcommand{\Mcal}{{\cal M}}
\newcommand{\Fcal}{{\cal F}}
\newcommand{\Ncal}{{\cal N}}

\newcommand{\gfrak}{{\mathfrak g}}

\newcommand{\ah}{{\hat{\smash{a}}}}
\newcommand{\bh}{{\hat{\smash{b}}}}

\def\im{\textrm{i}}

\def\diff{\textrm{d}}
\def\pa{\mbox{$\partial$}}
\def\sfrac#1#2{{\textstyle\frac{#1}{#2}}}
\def\+{\dagger}
\def\={\ =\ }

\def\and{\quad\textrm{and}\quad}
\def\with{\quad\textrm{with}\quad}
\def\for{\quad\textrm{for}\quad}

\def\Id{\mathrm{Id}}

\begin{document}

\begin{titlepage}
\setcounter{page}{0}
\begin{flushright}
ITP--UH--09/15
\end{flushright}

\hspace{2.0cm}

\begin{center}

{\Large\bf String theories as the adiabatic limit of Yang-Mills theory }

\vspace{12mm}

{\large  Alexander D. Popov}\\[8mm]

\noindent {\em
Institut f\"ur Theoretische Physik \\
Leibniz Universit\"at Hannover \\
Appelstra\ss e 2, 30167 Hannover, Germany }\\
{Email: popov@itp.uni-hannover.de
}\\[6mm]

\vspace{10mm}

\begin{abstract}
\noindent We consider Yang-Mills theory with a matrix gauge group $G$ on a direct product manifold $M=\Sigma_2\times H^2$,
where $\Sigma_2$ is a two-dimensional Lorentzian manifold and $H^2$ is a two-dimensional open disc with the boundary
$S^1=\partial H^2$. The Euler-Lagrange equations for the metric on $\Sigma_2$ yield  constraint equations for the Yang-Mills
energy-momentum tensor. We show that in the adiabatic limit, when the metric on $H^2$ is scaled down, the Yang-Mills equations
plus constraints on the energy-momentum tensor become the equations describing strings with a worldsheet $\Sigma_2$ moving in
the based loop group $\Omega G=C^\infty (S^1, G)/G$, where $S^1$ is the boundary of $H^2$. By choosing $G=\R^{d-1, 1}$ and
putting to zero all parameters in $\Omega\R^{d-1, 1}$ besides $\R^{d-1, 1}$, we get a string moving in $\R^{d-1, 1}$.
In arXiv:1506.02175 it was
described how one can obtain the Green-Schwarz superstring action from Yang-Mills theory on $\Sigma_2\times H^2$ while $H^2$
shrinks to a point. Here we also consider Yang-Mills theory on a three-dimensional manifold $\Sigma_2\times S^1$ and
show that
in the limit when the radius of $S^1$ tends to zero, the Yang-Mills action functional supplemented by a Wess-Zumino-type
term becomes the Green-Schwarz superstring action.
\end{abstract}

\end{center}
\end{titlepage}

\noindent {\bf 1. Introduction.} Superstring theory has a long history~\cite{1}-\cite{3} and pretends on description of all
four known forces in Nature. In the low-energy limit superstring theories describe supergravity in ten dimensions or
supergravity interacting with supersymmetric Yang-Mills (SYM) theory. On the other hand, Yang-Mills and SYM theories in four
dimensions give description of three main forces in  Nature not including gravity~\cite{4}-\cite{7}. The aim of this short
paper is to show that bosonic strings (both open and closed) as well as type I, IIA and IIB superstrings can be obtained as a subsector of pure Yang-Mills theory with
some constraints on the Yang-Mills energy-momentum tensor. Put differently, knowing the action for superstrings with a
worldsheet $\Sigma_2$, we introduce a Yang-Mills action functional on $\Sigma_2\times H^2$ or on $\Sigma_2\times S^1$
such that the Yang-Mills action becomes the Green-Schwarz superstring action while $H^2$ or $S^1$ shrink to a point.
We will work in Lorentzian signature, but all calculations can be repeated for the Euclidean signature of
spacetime.

\medskip

\noindent {\bf 2. Yang-Mills equations.}
 Consider Yang-Mills theory with a matrix gauge group $G$ on a direct product manifold $M=\Si_2\times H^2$, where
$\Si_2$ is a two-dimensional  Lorentzian manifold (flat case is included) with local coordinates $x^a, a,b,...=1,2$, and a
metric tensor $g^{}_{\Si_2}=(g_{ab})$, $H^2$ is the disc with coordinates $x^i$, $i,j,...= 3,4$, satisfying the inequality
$(x^3)^2+(x^4)^2<1$, and the metric $g_{H^2}=(g_{ij})$. Then  $(x^\mu )=(x^a, x^i)$ are local coordinates on $M$ with $\mu =
1,...,4$.

We start with the gauge potential $\Acal =\Acal_{\mu}\diff x^\mu$ with values in the Lie algebra $\gfrak=\mathrm{Lie}\, G$
having scalar product $(\cdot ,\cdot )$ defined either via trace Tr or, for abelian groups like $\R^{p, q}$,  $T^{p, q}$ etc.,
via a metric on vector spaces. The gauge field  $\Fcal =\diff\Acal + \Acal\wedge\Acal$ is the $\gfrak$-valued two-form
\begin{equation}\label{1}
 \Fcal =\sfrac12\Fcal_{\mu\nu}\diff x^\mu \wedge \diff x^\nu\with \Fcal_{\mu\nu} =\partial_\mu\Acal_\nu - \partial_\nu\Acal_\mu
 + [\Acal_\mu , \Acal_\nu]\ .
\end{equation}
The Yang-Mills equations on $M$ with the metric
\begin{equation}\label{2}
 \diff s^2 = g_{\mu\nu}\diff x^\mu \diff x^\nu = g_{ab}\diff x^a \diff x^b + g_{ij}\diff x^i \diff x^j
\end{equation}
have the form
\begin{equation}\label{3}
D_\mu\Fcal^{\mu\nu}:=\frac{1}{\sqrt{|\det g|}}\,\partial_\mu(\sqrt{|\det g|}\,\Fcal^{\mu\nu}) + [\Acal_\mu ,
\Fcal^{\mu\nu}]=0\ ,
 \end{equation}
where $g=(g_{\mu\nu})$ and $\pa_\mu ={\pa}/{\pa x^\mu}$.

The equations (\ref{3}) follow from the standard Yang-Mills action on $M$
\begin{equation}\label{4}
S=\sfrac14\int_M \diff^4x\,\sqrt{|\det g}|\,(\Fcal_{\mu\nu}, \Fcal^{\mu\nu})\ ,
\end{equation}
where $(\cdot ,\cdot )$ is the scalar product on the Lie algebra $\gfrak$. Note that the metric $g^{}_{\Si_2}$ on $\Si_2$ is
not fixed and the Euler-Lagrange equations for $g^{}_{\Si_2}$ yield the constraint equations
\begin{equation}\label{5}
 T_{ab}= g^{\la\sigma} (\Fcal_{a\la}, \Fcal_{b\sigma})-\sfrac14 \, g_{ab}(\Fcal_{\mu\nu}, \Fcal^{\mu\nu})=0
 \end{equation}
for the Yang-Mills energy-momentum tensor $T_{\mu\nu}$, i.e. its components along $\Si_2$ are vanishing. Note that these
constraints can be satisfied for many gauge  configurations, e.g. for self-dual gauge fields not only $T_{ab}=0$ but even
$T_{\mu\nu}=0$.

\medskip

\noindent {\bf 3. Adiabatic limit.} On $M=\Si_2\times H^2$ we have the obvious splitting
\begin{equation}\label{6}
\Acal =\Acal_{\mu}\diff x^\mu= \Acal_{a}\diff x^a+\Acal_{i}\diff x^i\ ,
\end{equation}
\begin{equation}\label{7}
 \Fcal =\sfrac12\Fcal_{\mu\nu}\diff x^\mu \wedge \diff x^\nu =\sfrac12\Fcal_{ab}\diff x^a \wedge \diff x^b + \Fcal_{ai}\diff x^a \wedge \diff x^i
+\sfrac12\Fcal_{ij}\diff x^i \wedge \diff x^j\ ,
 \end{equation}
\begin{equation}\label{8}
T =T_{\mu\nu}\diff x^\mu\diff x^\nu =T_{ab}\diff x^a\diff x^b + 2T_{ai}\diff x^a\diff x^i
+T_{ij}\diff x^i \diff x^j\ .
 \end{equation}
By using the adiabatic approach in the form presented in \cite{8, 9}, we deform the metric  (\ref{2}) and introduce the metric
\begin{equation}\label{9}
 \diff s^2_\ve = g_{ab}\diff x^a \diff x^b + \ve^2g_{ij}\diff x^i \diff x^j\ ,
\end{equation}
where $\ve\in [0,1]$ is a real parameter. It is assumed that the fields $\Acal_\mu$ and $\Fcal_{\mu\nu}$ smoothly depend
in $\ve^2$, i.e. $\Acal_\mu = \Acal_\mu^{(0)}+\ve^2 \Acal_\mu^{(1)}+ ...$   and   $\Fcal_{\mu\nu} = \Fcal_{\mu\nu}^{(0)}+
\ve^2 \Fcal_{\mu\nu}^{(1)}+ ...\ $. Furthermore, we have $\det g_\ve =\ve^4\det (g_{ab}) \det (g_{ij})$ and
\begin{equation}\label{10}
 \Fcal^{ab}_\ve = g_\ve^{ac}g_\ve^{bd}\Fcal_{cd}= \Fcal^{ab}\ ,\quad \Fcal^{ai}_\ve = g_\ve^{ac}g_\ve^{ij}\Fcal_{cj}=
 \ve^{-2}\Fcal^{ai}\and
 \Fcal^{ij}_\ve = g_\ve^{ik}g_\ve^{jl}\Fcal_{kl}=\ve^{-4}\Fcal^{ij}\ ,
\end{equation}
where indices in $\Fcal^{\mu\nu}$ are raised by the non-deformed metric tensor $g^{\mu\nu}$.

For the deformed metric  (\ref{9}) the action functional  (\ref{4}) is changed to
\begin{equation}\label{11}
S_\ve=\sfrac14\int_M \diff^4x\,\sqrt{|\det g^{}_{\Si_2}|}\,\sqrt{\det g^{}_{H_2}}\,\left\{\ve^2(\Fcal_{ab}, \Fcal^{ab}) +
2(\Fcal_{ai}, \Fcal^{ai}) + \ve^{-2}(\Fcal_{ij}, \Fcal^{ij})\right\}\ .
\end{equation}
The term $\ve^{-2}(\Fcal_{ij}, \Fcal^{ij})$ in the Yang-Mills Lagrangian (\ref{11}) diverges when
 $\ve\to 0$. To avoid this we
impose the flatness condition
\begin{equation}\label{12}
 \Fcal_{ij}^{(0)}=0\quad\Rightarrow\quad\lim_{\ve\to 0} (\ve^{-1}\Fcal_{ij})=0
\end{equation}
on the components of the field tensor along $H^2$. Here $\Fcal_{ij}^{(0)}=0$ but $\Fcal_{ij}^{(1)}$ etc. in the
$\ve^2$-expansion must not be zero.
For the deformed metric (\ref{9}) the Yang-Mills equations have the form
\begin{equation}\label{13}
 \ve^2D_a\Fcal^{ab} + D_i\Fcal^{ib}=0\ ,
\end{equation}
\begin{equation}\label{14}
 \ve\,D_a\Fcal^{aj} + \ve^{-1}D_i\Fcal^{ij}=0\ .
\end{equation}
In the deformed metric (\ref{9}) the constraint equations (\ref{5}) become
\begin{equation}\label{15}
 T_{ab}^{\ve}{=}\ve^2\left\{g^{cd}(\Fcal_{ac},\Fcal_{bd}){-}\sfrac14 g_{ab}(\Fcal_{cd},\Fcal^{cd})\right\}{+}g^{ij} (\Fcal_{ai}, \Fcal_{bj}){-}\sfrac12
 g_{ab}(\Fcal_{ci}, \Fcal^{ci}){-}\sfrac14 \ve^{-2}g_{ab}(\Fcal_{ij},\Fcal^{ij}){=}0 .
\end{equation}
In the adiabatic  limit $\ve\to 0$ the Yang-Mills equations (\ref{13}) and (\ref{14}) become
\begin{equation}\label{16}
 D_i\Fcal^{ib}=0\ ,
\end{equation}
\begin{equation}\label{17}
 D_a\Fcal^{aj}=0\ ,
\end{equation}
since the $\ve^{-1}$-term vanishes due to (\ref{12}).  We also keep (\ref{17}) since it follows from the action
(\ref{11}) after taking the limit $\ve\to 0$. One can see that the constraint equations (\ref{15}) are nonsingular in
the limit $\ve\to 0$ also due to (\ref{12}):
\begin{equation}\label{18}
T_{ab}^0 = g^{ij} (\Fcal_{ai}, \Fcal_{bj})-\sfrac12
 g_{ab}(\Fcal_{ci}, \Fcal^{ci})=0\ .
\end{equation}
Note that for the adiabatic limit of instanton equations \cite{8, 9} the constraints (\ref{15}) disappear since the
energy-momentum tensor for self-dual and anti-self-dual gauge fields vanishes on any four-manifold $M$.

\medskip

\noindent {\bf 4. Flat connections.} Now we start to consider the flatness equation (\ref{12}), the equations (\ref{16}),
(\ref{17}) and the constraint equations (\ref{18}). From now on we will consider only zero modes in $\ve^2$-expansions
and equations on them. For simplicity of notation we will omit the index ``$(0)$" from all $\Acal^{(0)}$ and $\Fcal^{(0)}$
tensor components.
In the adiabatic approach it is assumed that all fields depend on
coordinates $x^a\in\Si_2$ only via moduli parameters $\p^\a(x^a), \a , \b =1,2,...$, appearing in the solutions of the
flatness equation (\ref{12}).

Flat connection $\Acal_{H^2}:=\Acal_i\diff x^i$ on $H^2$ has the form
\begin{equation}\label{19}
 \Acal_{H^2}=g^{-1}\hat\diff g\with \hat\diff =\diff x^i\pa_i\for\pa_i=\frac{\pa}{\pa x^i}\ ,
\end{equation}
where $g=g(\p^\a(x^a), x^i)$ is a smooth map from $H^2$ into the gauge group $G$ for any fixed $x^a\in\Si_2$.

Let us introduce on $H^2$ spherical coordinates: $x^3=\rho\cos\vp$ and $x^4=\rho\sin\vp$. Using these coordinates, we  impose
on $g$ the condition $g(\vp =0, \rho^2\to 1)=\Id$ (framing) and denote by $C_0^\infty (H^2, G)$ the space of framed flat
connections on $H^2$ given by (\ref{19}). On $H^2$, as on a manifold with a boundary, the group of gauge transformations for
any fixed $x^a\in\Si_2$ is defined as (see e.g. \cite{9, 10, 11})
\begin{equation}\label{20}
 \Gcal_{H^2}= \left\{g: H^2\to G\mid g\to\Id\for \rho^2\to 1\right\}\ .
\end{equation}
Hence the solution space of the equation (\ref{12})  is the infinite-dimensional group $\Ncal = C_0^\infty (H^2, G)$, and the
moduli space of solutions is the based loop group \cite{9, 10, 12}
\begin{equation}\label{22}
 \Mcal = C^\infty_0 (H^2, G)/  \Gcal_{H^2}=\O G\ .
\end{equation}
This space can also be represented as $\O G = LG/G$, where $LG=C^\infty (S^1, G)$ is the loop group with the circle $S^1=\pa
H^2$ parameterized by $e^{\im\vp}$.

 \medskip

\noindent {\bf 5. Moduli space.} On the group manifold (\ref{22}) we introduce local coordinates $\p^\a$ with $\a =1, 2, ...$
and recall that $\Acal_\mu$'s depend on $x^a\in\Si_2$ only via moduli parameters $\p^\a = \p^\a (x^a)$. Then moduli of gauge
fields define a map
\begin{equation}\label{23}
 \p : \Si_2\to\Mcal\with \p (x^a)=\{\p^\a(x^a)\}\ .
\end{equation}
 These maps are constrained by the equations (\ref{16}), (\ref{17}) and (\ref{18}).
Since  $\Acal_{H^2}$ is a flat connection for any $x^a\in\Si_2$, the derivatives $\pa_a\Acal_i$ have to satisfy the linearized
(around  $\Acal_{H^2}$) flatness condition, i.e. $\pa_a\Acal_i$ belong to the tangent space $T_\Acal\Ncal$ of the space
$\Ncal=C^\infty_0 (H^2, G)$ of framed flat connections on $H^2$.
Using the projection $\pi : \Ncal\to\Mcal$ from  $\Ncal$ to the moduli space $\Mcal$, one can decompose $\pa_a\Acal_i$ into the two parts
\begin{equation}\label{24}
T_\Acal\Ncal= \pi^*T_\Acal\Mcal\oplus T_\Acal \Gcal \quad\Leftrightarrow\quad\pa_a\Acal_i=(\pa_a\p^\b)\xi_{\b i} + D_i\eps_a \ ,
\end{equation}
where $\Gcal$ is the gauge group $\Gcal_{H^2}$  for any fixed $x^a\in\Si_2$, $\{\xi_{\a}=\xi_{\a i}\diff x^i\}$ is a local
basis of tangent vectors at $T_\Acal\Mcal$ (they form the loop Lie algebra $\O\gfrak$) and $\eps_a$ are $\gfrak$-valued gauge
parameters ($D_i\eps_a\in T_\Acal \Gcal$) which are determined by the gauge-fixing conditions
\begin{equation}\label{25}
  g^{ij} D_i\xi_{\a j}=0\quad\Leftrightarrow\quad g^{ij} D_iD_j\eps_a=g^{ij} D_i \pa_a\Acal_j \ .
\end{equation}
Note also that since $\Acal_i(\p^\a , x^j)$ depends on $x^a$ only via $\p^\a$, we have
\begin{equation}\label{26}
 \pa_a\Acal_i=\frac{ \pa\Acal_i}{ \pa\phi^\b} \pa_a\phi^\b\quad\stackrel{\mathrm{(24)}}{\Longrightarrow}\quad \eps_a=(\pa_a\phi^\b)\,\eps_\b\ ,
\end{equation}
where the gauge parameters $\eps_\b$ are found by solving the equations
\begin{equation}\label{27}
g^{ij} D_i D_j\eps_\b=g^{ij} D_i\frac{ \pa\Acal_j}{ \pa\p^\b}\ .
\end{equation}

Recall that $\Acal_i$ are given explicitly by (\ref{19}) and  $\Acal_a$ are yet free. It is natural to choose $\Acal_a=\eps_a$ \cite{5, 6} and obtain
\begin{equation}\label{28}
 \Fcal_{ai}=\pa_a\Acal_i - D_i\Acal_a = (\pa_a\p^\b)\xi_{\b i}= \pi_*\pa_a\Acal_i \in T_\Acal\Mcal\ .
\end{equation}
Thus if we know the dependence of  $\p^\a$ on $x^a$ then we can construct
\begin{equation}\label{29}
 (\Acal_\mu )= (\Acal_a, \Acal_i)=\left( (\pa_a\p^\b)\,\eps_\b\ ,\ g^{-1}(\p^\a , x^j)\pa_i g(\p^\b , x^k)\right)\ ,
\end{equation}
which are in fact the components $\Acal_\mu^{(0)}{=}\Acal_\mu (\ve{=}0)$.

\medskip

\noindent {\bf 6. Effective action.} For finding equations for $\p^\a (x^a)$ we substitute (\ref{28}) into (\ref{16}) and see
that (\ref{16}) are resolved due to (\ref{25}). Substituting (\ref{28}) into (\ref{17}), we obtain the equations
\begin{equation}\label{30}
\frac{1}{\sqrt{|\det g^{}_{\Si_2}|}}\pa_a\left(\sqrt{|\det g^{}_{\Si_2}|}\ g^{ab}\pa_b\p^\b\right)g^{ij}\xi_{\b j}+
g^{ab}g^{ij}(D_a\xi_{\b j})\pa_b\p^\b= 0\ .
\end{equation}
We should project (\ref{30}) on the moduli space $\Mcal =\Omega G$, metric $\mathbb{G}=(G_{\a\b})$ on which is defined as
\begin{equation}\label{31}
 G_{\a\b}= \langle \xi_\a , \xi_\b\rangle =  \int_{H^2} d\, \mbox{vol} \ g^{ij}(\xi_{\a i}, \xi_{\b j})\ .
\end{equation}
The projection is provided by multiplying  (\ref{30}) by $ \langle \xi_\a , \cdot\rangle$ (cf. e.g. \cite{13, 14}). We obtain
\begin{equation}\nonumber
\frac{1}{\sqrt{|\det g^{}_{\Si_2}|}}\pa_a\left(\sqrt{|\det g^{}_{\Si_2}|}\ g^{ab}\pa_b\p^\b\right)\langle \xi_\a ,
\xi_\b\rangle + g^{ab}\langle \xi_\a , D_a\xi_\b\rangle\pa_b\p^\b
\end{equation}
\begin{equation}\label{32}
= \frac{1}{\sqrt{|\det g^{}_{\Si_2}|}}\pa_a\left(\sqrt{|\det g^{}_{\Si_2}|}\ g^{ab}\pa_b\p^\b\right)G_{\a\b} + \langle \xi_\a
, \nabla_\gamma\xi_\b\rangle g^{ab}\pa_a\p^\gamma \pa_b\p^\b
\end{equation}
\begin{equation}\nonumber
= G_{\a\sigma}\left\{\frac{1}{\sqrt{|\det g^{}_{\Si_2}|}}\pa_a\left(\sqrt{|\det g^{}_{\Si_2}|}\ g^{ab}\pa_b\p^\sigma\right) +
 \Gamma^\sigma_{\b\ga}g^{ab}\pa_a\p^\b\pa_b\p^\ga\right\} =0\ ,
\end{equation}
where
\begin{equation}\label{33}
 \Gamma^\sigma_{\b\ga}=\sfrac12\, G^{\sigma\la}\left(\pa_\ga \, G_{\b\la} +
 \pa_\b G_{\ga\la} - \pa_\lambda G_{\b\ga}\right)\with \pa_\ga :=\frac{\partial}{\partial \phi^\ga}\ ,
\end{equation}
are the Christoffel symbols and $\nabla_\ga$ are the corresponding covariant derivatives on the moduli space $\Mcal$ of flat
connections on $H^2$.

The equations
\begin{equation}\label{34}
\frac{1}{\sqrt{|\det g^{}_{\Si_2}|}}\pa_a\left(\sqrt{|\det g^{}_{\Si_2}|}\ g^{ab}\pa_b\p^\a\right) +
\Gamma^\a_{\b\ga}g^{ab}\pa_a\p^\b\pa_b\p^\ga =0
\end{equation}
are the Euler-Lagrange equations for the effective action
\begin{equation}\label{35}
 S_{\rm eff}=\int_{\Si_2} \diff x^1\diff x^2 \sqrt{-\det(g_{ab})}\, g^{cd}\, G_{\a\b}\, \pa_c\p^\a \pa_d\p^\b
\end{equation}
obtained from the action functional (\ref{11})  in the adiabatic limit $\ve\to 0$; it appears from the term $(\Fcal_{ai},
\Fcal^{ai})$ in (\ref{11})  (other terms vanish). The equations (\ref{34}) are the standard sigma-model equations defining
maps from $\Si_2$ into the based loop group $\Omega G$.

\medskip

\noindent {\bf 7. Virasoro constraints.} The last undiscussed equations are the constraints (\ref{18}). Substituting
(\ref{28}) into (\ref{18}), we obtain
\begin{equation}\label{36}
g^{ij}(\xi_{\a i},\xi_{\b j})\pa_a\p^\a \pa_b\p^\b - \sfrac12\, g_{ab}g^{cd}g^{ij}(\xi_{\a i},\xi_{\b j})\pa_c\p^\a \pa_d\p^\b =0\ .
\end{equation}
Integrating (\ref{36}) over $H^2$ (projection on $\Mcal$), we get
\begin{equation}\label{37}
 G_{\a\b}\pa_a\p^\a \pa_b\p^\b- \sfrac12\, g_{ab}g^{cd}G_{\a\b}\pa_c\p^\a \pa_d\p^\b =0\ .
\end{equation}
These are equations which one will obtain from (\ref{35}) by varying with respect to $g_{ab}$. Thus
\begin{equation}\label{38}
 T_{ab}^V=G_{\a\b}\pa_a\p^\a \pa_b\p^\b-\sfrac12\, g_{ab}g^{cd}G_{\a\b}\pa_c\p^\a \pa_d\p^\b
\end{equation}
is the traceless stress-energy tensor and equations  (\ref{37}) are the Virasoro constraints accompanying the Polyakov string action  (\ref{35}).

\medskip

\noindent {\bf 8. $B$-field.} In string theory the action (\ref{35}) is often extended by adding the $B$-field term. This term
can be obtained from the topological Yang-Mills term
\begin{equation}\label{39}
\sfrac12\,\int_M \diff^4x\,\sqrt{\det g^{\ve}_{H_2}}\, \ve_{\mu\nu\lambda\sigma} (\Fcal_{\ve}^{\mu\nu}, \Fcal^{\lambda\sigma}_\ve)
\end{equation}
which in the adiabatic limit $\ve\to 0$ becomes
\begin{equation}\label{40}
\int_M \diff^4x\,\sqrt{\det g_{H_2}}\, \ve^{ab} \ve^{ij}(\Fcal_{ai}, \Fcal_{bj})=\int_{\Si_2} \diff x^1\diff x^2 \,
\ve^{cd}\, B_{\a\b}\, \pa_c\p^\a \pa_d\p^\b\ ,
\end{equation}
where
\begin{equation}\label{41}
 B_{\a\b}= \int_{H^2} d\, \mbox{vol} \ \ve^{ij}(\xi_{\a i}, \xi_{\b j})\ .
\end{equation}
are components of the two-form $\mathbb{B}=(B_{\a\b})$ on the moduli space $\Mcal =\O G$.

\medskip

\noindent {\bf 9. Remarks on superstrings.} The adiabatic limit of supersymmetric Yang-Mills theories with a (partial)
topological twisting on Euclidean manifold $\Si\times \widetilde{\Si}$, where $\Si$ and $\widetilde{\Si}$ are Riemann
surfaces, was considered in \cite{15}. Several sigma-models with fermions on $\Si$ (including supersymmetric ones)
were obtained. Switching to Lorentzian
signature and adding constraints of type (\ref{18}), which were not considered in \cite{15}, one can get stringy sigma-model
resembling NSR strings. However, analysis of these sigma-models demands more efforts and goes beyond the scope of our paper.

Another possibility is to consider ordinary Yang-Mills theory (\ref{11}) but with Lie supergroup $G$ as the structure group.
We restrict ourselves to the $N{=}2$ super translation group with ten-dimensional Minkowski space $\R^{9,1}$ as bosonic part.
This super translation group can be represented as  the coset \cite{16,17}
\begin{equation}\label{42}
G=\mbox{SUSY($N{=}2$)/SO}(9,1)\ ,
\end{equation}
with coordinates $(X^\a, \th^{Ap})$, where $\th^p = (\th^{Ap})$ are two Majorana-Weyl spinors in $d=10, \a=0,...,9,
A=1,...,32$ and $p=1,2$.
The generators of $G$ obey the Lie superalgebra $\gfrak =\,$Lie$\,G$,
\begin{equation}\label{1a}
\{\xi_{Ap}, \xi_{Bq}\} =(\ga^\a C)_{AB}\de_{pq}\xi_\a\ ,\quad [\xi_{\a}, \xi_{Ap}]=0\ ,\quad [\xi_{\a}, \xi_{\b}]=0\ ,
\end{equation}
where $\ga^\a$ are the $\ga$-matrices in $\R^{9,1}$ and $C$ is the charge conjugation matrix. On the superalgebra
$\gfrak$ we introduce the standard metric
\begin{equation}\label{2a}
\langle\xi_{\a}\, \xi_{\b}\rangle=\eta_{\a\b}\ , \quad \langle\xi_{\a}\, \xi_{Ap}\rangle=0\and \langle\xi_{Ap}\, \xi_{Bq}\rangle=0\ ,
\end{equation}
where $(\eta_{\a\b})=\,$diag$(-1,1,...,1)$ is the Lorentzian metric on $\R^{9,1}$.

It was shown in \cite{18} that the action functional for Yang-Mills theory on $\Sigma_2\times H^2$ with the gauge group $G$,
defined by (\ref{1a}),
\begin{equation}\label{3a}
S_\ve=\frac{1}{2\pi}\,\int_{\Sigma_2\times H^2} \diff^4x\,\sqrt{|\det g^{}_{\Si_2}|}\,\sqrt{\det g^{}_{H_2}}\,\left\{\ve^2\langle\Fcal_{ab}\,
\Fcal^{ab}\rangle + 2\langle\Fcal_{ai}\, \Fcal^{ai}\rangle + \ve^{-2}\langle\Fcal_{ij}\, \Fcal^{ij}\rangle\right\}
\end{equation}
plus the Wess-Zumino-type term
\begin{equation}\label{4a}
 S_{WZ} = \frac{1}{\pi}\int_{\Si_3\times H^2}\diff x^\ah\wedge \diff x^\bh\wedge \diff x^{\hat c}\wedge\diff x^3\wedge \diff x^4\,
 f^{}_{\Gamma\Delta\Lambda}\, \Fcal^{\Gamma}_{\ah i}\xi^i \Fcal^{\Delta}_{\bh j}\xi^j\Fcal^{\Lambda}_{\hat c k}\xi^k
\end{equation}
yield the Green-Schwarz superstring action \cite{17} in the adiabatic limit $\ve\to 0$. Here $\Sigma_3$ is a Lorentzian manifold
with the boundary $\Sigma_2=\pa\Sigma_3$ and local coordinates $x^\ah$, $\ah =0,1,2$; the structure constants
$f^{}_{\Gamma\Delta\Lambda}$ are given in \cite{16} and $(\xi_i)=(\sin\varphi, -\cos\varphi)$ is the unit vector on
$H^2$ running the boundary $S^1=\pa H^2$.

\medskip

\noindent {\bf 10. Superstrings from $d=3$ Yang-Mills.} Here we will show that the Green-Schwarz superstrings
with a worldsheet $\Sigma_2$ can also be associated with a Yang-Mills model on $\Sigma_2\times S^1$. When the radius
of $S^1$ tends to zero, the action of this Yang-Mills model becomes the Green-Schwarz superstring action. So, we
consider Yang-Mills theory on a direct product manifold
$M^3= \Sigma_2\times S^1$, where $\Sigma_2$ is a two-dimensional Lorentzian manifold discussed before and $S^1$ is the
unit circle parameterized by $x^3\in [0,2\pi ]$ with the metric tensor $g^{}_{S^1}=(g_{33})$ and $g_{33}=1$. As the
structure group $G$ of Yang-Mills theory we consider the super translation group in $d=10$ auxiliary dimensions (\ref{42})
with the generators (\ref{1a}) and the metric (\ref{2a}) on the Lie superalgebra $\gfrak =\,$Lie$\,G$. As in (\ref{20}),
we impose framing over $S^1$, i.e. consider the group of gauge transformations equal to the identity over $S^1$.
Coordinates on $G$
are $X^\a$ and $\th^{Ap}$ introduced in the previous section. The one-forms
\begin{equation}\label{46}
\Pi^{\Delta}= \{\Pi^\a , \Pi^{Ap}\} =\{\diff X^\a-\im\,\delta_{pq}\,\bar\th^p\ga^\a\diff\th^q ,\ \diff\th^{Ap}\}
\end{equation}
form a basis of one-forms on $G$ \cite{16}.

By using the adiabatic approach, we deform the metric on $\Sigma_2\times S^1$ and introduce
\begin{equation}\label{47}
 \diff s^2_\ve = g_{\mu\nu}^\ve\,\diff x^\mu \diff x^\nu = g_{ab}\diff x^a \diff x^b + \ve^2(\diff x^3)^2\ ,
\end{equation}
where $\ve\in [0,1]$ is a real parameter, $a,b=1,2$, $\mu,\nu = 1,2,3.$ This is equivalent to the consideration of the circle
$S^1_\ve$ of radius $\ve$. It is assumed that for the fields $\Acal_\mu$ and $\Fcal_{\mu\nu}$ there exist limits
$\lim_{\ve\to 0}\Acal_\mu$ and $\lim_{\ve\to 0}\Fcal_{\mu\nu}$.
   Indices are raised by $g_\ve^{\mu\nu}$ and we have
\begin{equation}\label{48}
 \Fcal^{ab}_\ve = g_\ve^{ac}g_\ve^{bd}\Fcal_{cd}= \Fcal^{ab}\ ,\quad \Fcal^{a3}_\ve = g_\ve^{ac}g_\ve^{33}\Fcal_{c3}=
 \ve^{-2}\Fcal^{a3}\ ,
\end{equation}
where indices in $\Fcal^{\mu\nu}$ are raised by the non-deformed metric tensor.

We consider the Yang-Mills action of the form
\begin{equation}\label{49}
S_\ve=\int_{M^3} \diff^3x\,\sqrt{|\det g^{}_{\Si_2}|}\,\left\{\frac{\ve^2}{2}\langle\Fcal_{ab}\,
\Fcal^{ab}\rangle + \langle\Fcal_{a3}\, \Fcal^{a3}\rangle\right\}\ ,
\end{equation}
which for $\ve =1$ coincides with the standard Yang-Mills action. Variations with respect to $\Acal_\mu$ and $g_{ab}$
yield the equations
\begin{equation}\label{50}
\ve^2D_a\Fcal^{ab} + D_3\Fcal^{3b}=0\ ,\quad
 D_a\Fcal^{a3}=0\ ,
\end{equation}
\begin{equation}\label{51}
T^\ve_{ab}=\ve^2\left(g^{cd}\langle\Fcal_{ac}\, \Fcal^{bd}\rangle -
\sfrac14\,g_{ab}\langle\Fcal_{cd}\, \Fcal^{cd}\rangle\right)
+ \langle\Fcal_{a3}\, \Fcal_{b3}\rangle
-\sfrac12\,g_{ab}\langle\Fcal_{c3}\, \Fcal^{c3}\rangle \ .
\end{equation}
In the adiabatic limit $\ve\to 0$ equations (\ref{50}), (\ref{51}) become
\begin{equation}\label{52}
D_3\Fcal^{3b}=0\ ,\quad
 D_a\Fcal^{a3}=0\ ,
\end{equation}
\begin{equation}\label{53}
T^0_{ab}=\langle\Fcal_{a3}\, \Fcal_{b3}\rangle
-\sfrac12\,g_{ab}\langle\Fcal_{c3}\, \Fcal^{c3}\rangle \ .
\end{equation}

Notice that as a function of $x^3\in S^1$, the field $\Acal_3$ belongs to the loop algebra $L\gfrak =
\gfrak\oplus\Omega\gfrak$, where $\Omega\gfrak$ is the Lie superalgebra of the based loop group $\Omega G$.
Let us denote by $\Acal^0_3$ the zero-mode in the expansion of $\Acal_3$ in $\exp (\im\,x^3)\in S^1$
(Wilson line).  The generic $\Acal_3$ can be represented in  the form
\begin{equation}\label{54}
\Acal_3=h^{-1}\Acal^0_3 h + h^{-1}\pa_3 h \ ,
\end{equation}
where $G$-valued function $h$ depends on $x^a$ and $x^3$. For fixed $x^a\in\Si_2$ one can choose
$h\in\Omega G$=Map$(S^1, G)/G$.
We denote by $\Ncal$  the space of all $\Acal_3$ given by (\ref{54}) and define the projection $\pi : \Ncal\to G$
on the space $G$ parametrizing $\Acal_3^0$ since we want to keep only $\Acal_3^0$ in the limit
$\ve\to 0$. We denote by $Q$ the fibres of the projection $\pi$.

In the adiabatic approach it is assumed that $\Acal_3^0$ depends on $x^a\in\Si_2$ only via
the moduli parameters $(X^\a , \th^{Ap})\in G$. Therefore, the moduli define the maps
\begin{equation}\label{54a}
(X, \th^p ):\quad \Si_3\to G
\end{equation}
which are not arbitrary, they are constrained by the equations (\ref{52}), (\ref{53}). The derivatives
$\pa_a\Acal_3$ of $\Acal_3\in \Ncal$ belong to the tangent space $T_{\Acal_3}\Ncal$ of the space $\Ncal$.
Using the projection $\pi :\Ncal\to G$, one can decompose $\pa_a\Acal_3$ into two parts
\begin{equation}\label{55}
T_{\Acal_3}\Ncal= \pi^*T_{\Acal^0_3} G\oplus T_{\Acal_3}Q \quad\Leftrightarrow\quad
\pa_a\Acal_3=\Pi_a^{\Delta}\xi_{\Delta 3} + D_3\eps_a \ ,
\end{equation}
where $\Delta = (\a , Ap)$ and
\begin{equation}\label{56}
\Pi_a^\a :=\pa_a X^\a - \im\,\delta_{pq}\,\bar\th^p\ga^\a\pa_a\th^q\ ,\quad \Pi_a^{Ap}:=\pa_a\th^{Ap}\ .
\end{equation}
In (\ref{55}), $\eps_a$ are $\gfrak$-valued parameters ($D_3\eps_a\in T_{\Acal_3}Q$) and the vector fields
 $\xi_{\Delta 3}$ on $G$ can be identified  with the generators $\xi_\Delta =(\xi_\a, \xi_{Ap})$ of $G$.

On $\xi^{}_{\Delta 3}$  we impose the gauge fixing condition
\begin{equation}\label{57}
 D_3\xi^{}_{\Delta 3}=0\quad\stackrel{\mathrm{(56)}}{\Longrightarrow}\quad D_3D_3\eps_a=D_3 \pa_a\Acal_3 \ .
\end{equation}
Recall that $\Acal_3$ is fixed by (\ref{54}) and $\Acal_a$ are yet free. In the adiabatic approach one chooses
$\Acal_a=\eps_a$ (cf. \cite{5,6}) and obtains
\begin{equation}\label{58}
\Fcal_{a3}= \pa_a\Acal_3 - D_3\Acal_a= \Pi_a^{\Delta}\xi_{\Delta 3} \in T^{}_{\Acal_3^0}G\ .
\end{equation}
Substituting (\ref{58}) into the first equation in (\ref{52}), we see that they are resolved due to (\ref{57}).
Substituting (\ref{58}) into the action $S_0=\lim_{\ve\to 0} S_\ve$ given by (\ref{49}) and integrating over $x^3$,
we obtain the effective action
\begin{equation}\label{59}
S_0=2\pi\int_{\Si_2} \diff^2 x\,\sqrt{|\det g^{}_{\Si_2}|}\, g^{ab}\,\Pi_a^\a\,\Pi_b^\b\,\eta_{\a\b}\ ,
\end{equation}
which coincides with the kinetic part of the Green-Schwarz superstring action \cite{17}. One can show (cf. \cite{14})
that the second equations in (\ref{52}) are equivalent to the Euler-Lagrange equations for $(X^\a , \th^{Ap})$
following from (\ref{59}). Finally, substituting (\ref{58}) into (\ref{53}), we obtain the equations
\begin{equation}\label{60}
 \Pi_a^\a\,\Pi_b^\b\,\eta_{\a\b}  - \sfrac12\, g_{ab}\, g^{cd} \,\Pi_c^\a\,\Pi_d^\b \,\eta_{\a\b} =0
\end{equation}
which can also be obtained from (\ref{59}) by variation of $g^{ab}$.

For getting the full Green-Schwaz superstring action one should add to (\ref{59}) a Wess-Zumino-type term  which is
described as follows \cite{16,17}. One should consider a Lorentzian 3-manifold $\Si_3$ with the boundary $\Si_2=\pa\Si_3$
and coordinates $x^\ah$, $\ah = 0,1,2$. On $\Si_3$ one introduces the 3-form \cite{16}
\begin{equation}\label{61}
\O_3= \im\,\diff x^\ah\Pi_\ah^\a\wedge(\check\diff\bar\th^1\ga^\b\wedge\check\diff\th^1 -
\check\diff\bar\th^2\ga^\b\wedge\check\diff\th^2) \,\eta_{\a\b}=\check\diff\O_2\ ,
\end{equation}
where
\begin{equation}\nonumber
\O_2=-\im\check\diff X^\a\wedge(\bar\th^1\ga^\b\check\diff\th^1 - \bar\th^2\ga^\b\check\diff\th^2)\with
\check\diff= \diff x^\ah\frac{\pa}{\pa x^\ah}\ .
\end{equation}
Then the term
\begin{equation}\label{62}
 S_{WZ} = \int_{\Si_3}\O_3 = \int_{\Si_2}\O_2
\end{equation}
is added to (\ref{59}) with a proper coefficient $\vk$ and $S_{GS}= S_{0} +\vk S_{W\!Z}$ is the Green-Schwarz action
for the superstrings of type I, IIA and IIB.

To get (\ref{62}) from Yang-Mills theory we consider the manifold $\Si_3\times S^1$ and notice that in addition to
(\ref{58}) we now have the components
\begin{equation}\label{63}
\Fcal_{03}=\Pi_0^\Delta\xi_{\Delta 3}=(\pa_0X^\a - \im\,\de_{pq}\bar\th^p\ga^\a\pa_0\th^q)\xi_{\a 3}+
(\pa_0\th^{Ap})\xi_{Ap3} .
 \end{equation}
Introduce one-forms $F_3:=\Fcal_{\ah 3}\diff x^\ah$ on $\Si_3$, where $\Fcal_{\ah 3}(\ve)$ are general Yang-Mills fields on
$\Si_3\times S^1$ which take the form (\ref{58}),(\ref{63}) only in the limit $\ve\to 0$, and consider the functional
\begin{equation}\label{64}
 S_{WZ}^{\ YM} = \int_{\Si_3\times S^1}\,f^{}_{\Delta\Lambda\Gamma}\,F_3^\Delta\wedge F_3^\Lambda\wedge F_3^\Gamma\wedge\diff x^3 \ ,
\end{equation}
where the explicit form of the constant $f^{}_{\Delta\Lambda\Gamma}$ can be found in \cite{16}. Therefore, the
Yang-Mills action
(\ref{49}) plus (\ref{64}) in the adiabatic limit $\ve\to 0$ becomes the Green-Schwarz action. This result
can be considered
as a generalization of the Green result \cite{19} who derived the superstring theory in a fixed gauge from
Chern-Simons theory on $\Si_2\times\R$.

\medskip

\noindent {\bf 11. Concluding remarks.} We have shown that bosonic strings and Green-Schwarz superstrings can be
obtained via the adiabatic limit of Yang-Mills theory on manifolds $\Si_2\times H^2$ with a Wess-Zumino-type term.
Notice that
the constraint equations (\ref{15}) on the Yang-Mills energy momentum tensor with $\ve >0$ are important for restoring
the unitarity of Yang-Mills theory on $\Si_2\times H^2$. More interestingly, the same result is also obtained
by considering Yang-Mills theory on three-dimensional manifolds $\Si_2\times S^1_\ve$ with the radius of the
circle $S^1_\ve$
given by $\ve\in [0,1]$. For $\ve\ne 0$ we have well-defined quantum Yang-Mills theory on $\Si_2\times S^1_\ve$.
For $\ve\to 0$ we get superstring theories. This raises hopes that various results for superstring theories
can be obtained
from results of the associated Yang-Mills theory on $\Si_2\times S^1_\ve$.

\medskip

\noindent
{\bf Acknowledgements}

\noindent
This work was partially supported by the Deutsche Forschungsgemeinschaft grant LE 838/13.


\end{document}